\documentclass[12pt]{article}
\usepackage{graphics}
\usepackage{amsthm}
\usepackage{amssymb}
\usepackage{amsmath}
\usepackage{amsfonts}
\usepackage{epic}
\usepackage{hyperref}
\usepackage{epsfig}
\usepackage{bm}
\usepackage{amscd}

\theoremstyle{plain}

\theoremstyle{definition}

\def\be{\begin{equation}}
\def\ee{\end{equation}}
\smallskip

\begin{document}

\headsep=-0.5cm

\begin{titlepage}
\begin{flushright}
\end{flushright}
\begin{center}
\noindent{{\LARGE{Comments on the two-point string amplitudes}}}

\smallskip
\smallskip
\smallskip

\smallskip
\smallskip

\smallskip

\smallskip

\smallskip
\smallskip
\noindent{\large{Gaston Giribet, Nicholas Labranche, Joan La Madrid}}
\end{center}

\smallskip

\smallskip
\smallskip

\centerline{Department of Physics, New York University}
\centerline{{\it 726 Broadway, New York, NY10003, USA.}}

\smallskip
\smallskip

\smallskip
\smallskip

\smallskip

\smallskip

\begin{abstract}
The two-point string amplitude at tree level in flat spacetime reproduces the expected expression for free particles. This has been proven by Erbin, Maldacena and Skliros in [JHEP {\bf 07} (2019) 139] by two methods. Here, we provide an alternative proof of this result. Our method consists in considering a timelike Liouville direction as a regularization parameter, which suffices to break time translation invariance and, at the same time, to stabilize the residual conformal symmetry in the worldsheet. This allows taking a limit and restoring Poincar\'e symmetry in a controlled way, thus reproducing the correct expression of the two-point amplitude.

\end{abstract}

\end{titlepage}

\newpage




\section{Introduction}

In \cite{Malda}, Erbin, Maldacena and Skliros showed that the two-point string amplitude at tree level in flat spacetime reproduces the expected expression for free particles. Proving this amounts to show how to stabilize the residual symmetry that persists after fixing the two vertex operators on the projective sphere, or, more precisely, how the divergent denominator that comes from modding out such residual symmetry is cancelled by another divergence that comes from the integration over a zero mode in the non-compact target spacetime. In a two-point correlation function, one only has at hand two vertices to be fixed on the worldsheet, while three are needed at tree-level to cancel the volume of the non-compact conformal Killing group: this produces a $1/\infty $ factor. On the other hand, in the two-point string amplitude, since the states are on-shell, the momentum conservation directly implies the energy conservation, and this yields a singular factor $\delta (0)=\infty $. That is to say, the latter singularity follows from time-translation invariance. Interestingly, both infinities cancel each other producing the standard contribution to the $S$-matrix of a quantum field theory, as required by unitarity \cite{Malda}. This observation turns out to be important as the standard lore in the literature is that the string two-point amplitude on the sphere is zero \cite{DHP}: for example, in seminal papers we read that, if two vertex operators are inserted, then one should fix two points and it remains to divide out by the volume of the isotropy subgroup leaving two points invariant; and when one fixes these points at zero and infinity, the group is that of rotations and dilations and again has infinite volume, so that the two-point function vanishes to tree level\footnote{For interesting discussions about the 2-point string amplitudes see also \cite{BT1, BT2}.}
.

The finite, non-zero result for the two-point string amplitude was recently obtained in \cite{Malda} by means of two different methods: First, the authors considered a worldsheet classical solution that spontaneously breaks time-translation symmetry. The symmetry is of course restored once the integration over the zero-mode of the time direction ($x^0$) is performed, but the spontaneous symmetry breaking allows to have control on the ratio of divergent quantities. The second method considered in \cite{Malda} resorts to the Faddeev-Popov trick. The results of both methods are in complete agreement and yield a sensible finite answer. This was later confirmed by other computations: in \cite{190903672} the authors considered the operator formalism, and in \cite{2012.03802} a calculation using the pure spinor formalism was addressed. Here, we will present another, qualitatively different method to obtain the tree-level two-point string amplitude: we will consider an explicit symmetry breaking of time-translation by introducing worldsheet operators that, while preserving conformal invariance in the worldsheet, break Poincar\'e invariance in the target space. This amounts to deal with a solvable time-dependent background in string theory, which is achieved by considering a timelike Liouville direction that allows to take a limit in which time-translation symmetry is eventually restored in a controlled way. The paper is organized as follows: In section 2, we will discuss the two-string amplitudes at tree-level and the problem one is dealing with. In section 3, we will introduce the time-like Liouville background that will work as a regulator to compute the two-point amplitude. We will discuss in details each step in the calculation that leads to the correct result. Section 4 contains some concluding remarks on the reflection coefficient and the correlation functions in the timelike Liouville theory.

\section{Two-point string amplitudes}

Let us start by considering the Polyakov action
\begin{equation}
S_P=\frac{1}{2 \pi \alpha'}\int_{\Sigma } d^{2}z\left( \delta _{ij}\, \partial X^i \bar{\partial }X^j -\partial X^0 \bar{\partial }X^0  \right) \, ,  \label{manchaa}
\end{equation}%
with $i, j\in \{1, 2, ... , D-1\}$. This action describes bosonic string theory in $D$ spacetime dimensions. A generalization of our discussion to backgrounds of the form $\mathbb{R}^{1,D-1}\times \mathcal{M}$, with $\mathcal{M}$ being an internal compact manifold is straightforward, as it is to consider the superstring extension; therefore, (\ref{manchaa}) suffices to make our point. 

The $N$-point string amplitudes are given by the correlation numbers\footnote{
For short, we will omit normal ordering symbols $:\, :$ , as well as explicit reference to ghost contributions $\langle c\bar{c}{(z_1)} \, ... \, c\bar{c}{(z_3)}\rangle _{\text{ghosts}}$.} 
\begin{equation}
\mathcal{A}_N(p^{\mu}_{(1)}, p^{\mu}_{(2)}, ..., p^{\mu}_{(N)})=\langle V_{p_{(1)}}\, V_{p_{(2)}}\, ...\, V_{p_{(N)}}\rangle\label{tarde2}
\end{equation}
which correspond to integrated $N$-point correlation functions of dimension-($1,1$) primary operators $V_{p_{(a)}}$ in the worldsheet CFT defined by (\ref{manchaa}). The label $p_{(a)}$ is the $D$-momentum of the $a^{\text{th}}$ vertex ($a=1, ... , N$) with components $p^{\mu }_{(a)}=(p^0_{(a)}, p^i_{(a)})$ ($\mu = 0, 1, ..., D-1$ and $i = 1, ..., D-1$). The expectation value in (\ref{tarde2}) is defined on the Riemann surface that describes the worldsheet ${\Sigma }$; here, ${\Sigma }=S^2$. For example,  consider the illustrative example of tachyon $N$-point amplitudes at tree-level, which are defined by integrated correlators of exponential operators on the Riemann sphere; namely
\begin{equation}
\langle V_{p_{(1)}}\, V_{p_{(2)}}\, ...\, V_{p_{(N)}}\rangle = \int _{\mathbb{C}^{N}} \prod_{a=1}^Nd^2z_a\, \int_{X^{\mu }_{(\mathbb{CP}^1)}} \frac{\mathcal {D}X^{\mu }\, e^{-S_P}}{\text{Vol}(PSL(2,\mathbb{C}))} \, \prod_{b=1}^N \, e^{ip^{(b)}_{\mu_{b}}X^{\mu_{b}}(z_b)}\label{tarde3}
\end{equation}
with $a,b=1, 2, ..., N$ and $\mu , \mu_b = 0,1, ...., D-1$. At tree-level, we have ${\Sigma }=S^2$ with $N$ punctures and thus the path integral is defined on configurations $X^{\mu}(z)$ with suitable boundary conditions on $\mathbb{CP}^1\backslash \{ z_1, ..., z_N\} $. The factor $\text{Vol}_{(PSL(2,\mathbb{C}))}$ in the denominator of (\ref{tarde3}) is the stabilizer of the conformal symmetry, which fixes the gauge ambiguity under projective invariance given by the M\"obius group $PSL(2,\mathbb{C})=SL(2,\mathbb{C})/\mathbb{Z}_2$: Namely, this is the volume of the non-compact conformal Killing group on the projective sphere. The subtle point in the computation of the two-point function is precisely taking care of such volume. The non-compactness of the conformal Killing group makes the two-point functions per unit of target space volume to be zero. The usual way of dealing with the factor $\text{Vol}^{-1}_{(PSL(2,\mathbb{C}))}$ in the $N$-point amplitude with $N\geq 3$ is fixing 3 out of $N$ insertion points of the vertices at arbitrary points of $\mathbb{CP}^1$, usually taken to be $z_1=0, z_2=1$ and $z_N=\infty $. This cancels the volume factor of the group $PSL(2,\mathbb{C})$; see (\ref{quince}) below. Therefore, for the $N$-point amplitude with $N\geq 3$ the calculation reduces to integrating over $N-3$ points of the worldsheet $N$-point correlation functions on the Riemann sphere. However, there exists an obviously obstruction in the case $N=2$, as a residual symmetry remains after the two vertices are fixed and there is still a divergent factor in the denominator. On the other hand, in the case of target space with Poincar\'e invariance, there is an extra infinite factor in the numerator of the amplitude: since the string states are on-shell, in the case of the two-point amplitude the $(D-1)$-dimensional $\delta $-function coming from the integration over the zero-modes of the spacelike fields $X^i(z)$, which realizes the momentum conservation, automatically implies energy conservation, so producing an already-evaluated factor $\delta (0)$. This yields an indeterminacy of the form
\begin{equation}
\mathcal{A}_2(p^{\mu}_{(1)},p^{\mu}_{(2)})= \frac{\infty }{\infty } \, ,\label{A2}
\end{equation}
which is necessary to resolve. Resolving it demands to take control over both divergent quantities. Here, we will do this by introducing a Liouville dimension as a regulator, which will provide us with a method to restore time translation symmetry in a controlled way. 

One can anticipate the final result, cf. \cite{Malda}. Since the string two-point amplitude is expected to reproduce the standard free particle expression, which satisfies the requirement
\begin{equation}
\mathcal{A}_2(p^{\mu}_{(1)},p^{\mu}_{(2)}) = \int_{\mathbb{R}^{D}}\frac{d^Dq}{(2\pi )^{D-1}}\,
\,\Theta (q^0)\, \delta (q_{\nu } q^{\nu }+M^2)\, \mathcal{A}_2(p^{\mu}_{(1)},q^{\mu})\, \mathcal{A}_2(q^{\mu},p^{\mu}_{(2)}) \,\label{ecua} 
\end{equation}
with the mass-shell condition $M^2=-q_{\mu }q^{\mu }=(q^0)^2-q_iq^i$, then the two-point amplitude, provided it is finite and non-zero, has to be
\begin{equation}
\mathcal{A}_2(p^{\mu}_{(1)},p^{\mu}_{(2)}) =   2p_{(1)}^0\, (2\pi )^{D-1}\delta^{(D-1)} (p^i_{(1)}+ p^i_{(2)})\, .\label{SP2}
\end{equation}
As said in the Introduction, showing that (\ref{A2}) actually yields (\ref{SP2}) demands taking care of the two competing divergences in the former expression and taking the limit in a controlled way. This is exactly what the authors of \cite{Malda} did. Here, we will consider a rather different approach: we will break Poincar\'e invariance explicitly by adding exact marginal operators in the worldsheet CFT. While preserving conformal invariance, such operators will suffice to break the shift symmetry under time-translation $X^0(z)\to X^0(z)+x^0$ in a controlled way. This actually means to find a solvable time-dependent background in string theory. This can be done by turning $X^0(z)$ into a timelike Liouville direction. This gives a solvable, time-dependent dilaton-tachyon deformation of the flat background, which permits to restore the Poincar\'e symmetry in the appropriate limit, where the deformation vanishes.

\section{Liouville regularization}

\subsection{Time-dependent background}

We will consider a time dependent deformation of (\ref{manchaa}). This consists in turning the time direction $X^0$ into a timelike Liouville direction. In such case the worldsheet CFT reduces to timelike Liouville field theory coupled to $D-1$ spacelike free scalars $X^i$ ($i=1, 2, ... , D-1$); namely\footnote{We set $\alpha '=2$ hereafter.}
\begin{equation}
S[\mu ]=\frac{1}{4\pi }\int_{\Sigma } d^{2}z\left(  \delta _{ij}\, \partial X^i \bar{\partial }X^j -\partial X^0 \bar{%
\partial }X^0 +\frac{1}{\sqrt{2}}QRX^0 +4\pi \mu e^{\sqrt{2}%
bX^0 } \right) \, ,  \label{mancha}
\end{equation}%
with 
\begin{equation}
Q=b-\frac{1}{b}\, .\label{ElQ}
\end{equation}
This admits an interpretation as two-dimensional quantum gravity coupled to free matter fields. From the string theory point of view, this corresponds to consider a time-dependent background with the following dilatonic and tachyonic configurations
\begin{equation}
\Phi (X) = \frac{Q}{2\sqrt 2}\, X^0 +\Phi_0 \, ,  \ \ \ \ T(X) = 2\pi\mu \, e^{\sqrt 2 b X^0} \, , 
\end{equation}
respectively. $\mu $ is a positive parameter that controls the strength of the tachyon wall potential. Its value can be changed as $\mu \to \mu_*$ by shifting
\begin{equation}
X^0 (z) \to X^0 (z) +\frac{1}{\sqrt{2}b} \log \left( \frac{\mu_*}{\mu}\right)\, ,\label{Diego}
\end{equation}
which is equivalent to shifting the dilaton expectation value $\Phi_0 = \langle \Phi \rangle $, which means rescaling the string coupling. This is associated to the fact that the value of $\mu $ only enters as an overall factor in the Liouville observables, controlling the Knizhnik-Polyakov-Zamolodchikov (KPZ) scaling of the worldsheet correlators. Keeping track of the $\mu $-dependence will be important because, whatever limit of (\ref{mancha}) we could be interested in, this has to be consistent with keeping the string coupling finite. 

The background charge $Q$ in (\ref{mancha}) takes the value (\ref{ElQ}) for the Liouville potential barrier $\mu e^{\sqrt{2}bX^0 }$ to be an exact marginal operator. In other words, this ensures the time-dependent deformation to preserve conformal invariance. -- There is a second marginal operator, $ e^{-\sqrt{2}X^0/b }$, which, in the timelike theory, has the opposite sign in the exponent--. The third term in (\ref{mancha}), which is the dilaton term, involves the scalar curvature $R$, which in the conformal gauge has to be understood as keeping record of a $\delta$-function contribution to the $S^2$ curvature coming from the point at
infinity in $\mathbb{CP}^1$.

The central charge of the theory defined by action (\ref{mancha}) is
\begin{equation}
c=D-6Q^2=D-6(b-1/b)^{2},\label{tardec}
\end{equation}
which, provided $b\in \mathbb{R}$, lies in the semi-infinite segment $- \infty <c\leq D$ (with the Liouville contribution to the central charge being $c_L=c-D+1=1-6Q^2$). Notice that we could have considered a more general background of the form $\mathbb{R}^{1,D-1}\times \mathcal{M}$, and in that case (\ref{tardec}) would have received an extra term $c_{\mathcal{M}}$. We take $\mathcal{M}=T^{D-26}$ for simplicity.

Here, we are interested in the Liouville operator only as a regularization trick that will allow us to obtain a finite result for the flat space two-point amplitude. In other words, we will compute the two-point function in the time-dependent background and, after that, we will proceed to turn off the Liouville deformation and restore time translation symmetry. More precisely, the specific limit we want to take is
\begin{equation}
b=1+\varepsilon \to 1 \, , \ \ \ \ Q\simeq 2\varepsilon \to 0 \, , \label{LaLimite}
\end{equation}
and, at the same time,
\begin{equation}
\mu \sim \epsilon \to 0\, .\label{LaLimite2}
\end{equation}
The latter amounts to keep the string coupling finite while turning off the Liouville deformation. We find convenient to take $\pi \mu =- Q$. In the limit (\ref{LaLimite})-(\ref{LaLimite2}), taking into account the ghosts contribution, $c_{\text{ghosts}}=-26$, the critical condition yields
\begin{equation}
D\simeq 26+24\varepsilon ^2\, ,
\end{equation}
which can be thought of as a sort of dimensional regularization in target space. Target space dimensional regularization using the Coulomb gas approach has been considered in the context of superstring amplitudes in \cite{Ishibashi}.

\subsection{Liouville two-point function}

The two-point tachyon amplitude in the theory defined by (\ref{mancha}) is given by 
\begin{equation}
\mathcal{A}_2 (p^{\mu}_{(1)}, p^{\mu}_{(2)})=\int _{X^{\mu}_{(\mathbb{CP}^1)} }\frac{\mathcal{D}X^0 \mathcal{D}X^i \, \, e^{-S[\mu ]}}{\text{Vol}{(Res)}} \, e^{i\sqrt{2}p^{(1)}_{i_1}X^{i_1 }(0)+\sqrt{2}\alpha X^0(0)} \,e^{i\sqrt{2}p^{(2)}_{i_2}X^{i_2 }(1)+\sqrt{2}\alpha X^0(1)}\, ,\label{Defa}
\end{equation}%
where we have fixed $z_1=0$ and $z_2=1$, and where the Liouville momenta are taken to be $\alpha =Q/2+ip^0_a$; here, $a = 1, 2$ while $i, i_a =1, ... , D-1$ and $\mu  =0, 1, ... , D-1$. These values of $\alpha $  are the natural generalization of the momentum of normalizable states in spacelike ($c\geq 25$) Liouville field theory. 

The non-trivial part of the calculation comes from the Liouville piece, which is the non-gaussian contribution to the worldsheet theory. Due to the exponential form of the vertices $V_{p_{(a)}}$, the integral over the zero modes ($x^i$) of the spacelike fields $X^i(z)$ ($i=1,..., D-1$) produces a momentum $\delta $-function. --For other string states, such as massless states, the same computation holds.-- This means that we can separate that dependence as follows
\begin{equation}
\mathcal{A}_2 (p^{\mu}_{(1)}, p^{\mu}_{(2)}) = (2\pi )^{D-1}\delta^{D-1} (p^i_{(1)}+ p^i_{(2)})\, \lim_{\varepsilon \to 0 }\,\mathcal{A}_2' (p^0_{(1)}) 
\end{equation}
where $\mathcal{A}_2' (p^0)$ stands for the Liouville factor
\begin{equation}
\mathcal{A}_2' (p_1^0)=\int _{X^0_{(\mathbb{CP}^1)}} \frac{\mathcal{D}X^0 \, \, e^{-S_L[\mu ]}}{\text{Vol}{(Res)}} \, e^{\sqrt{2}\alpha X^0(0)}\, e^{\sqrt{2}\alpha X^0(1)} \, ,\label{Defa}
\end{equation}%
with the timelike Liouville action $S_L[\mu ]=S[\mu ]-\frac{1}{4\pi}\int_{\Sigma } d^2z\, \partial X_i \bar{\partial} X^i$. Therefore, the problem reduces to that of solving the timelike Liouville two-point function (\ref{Defa}) in the limit of the Liouville central charge $c_L\to 1$. This limit is known to be subtle, especially in what regards the three-point function, cf. \cite{S}; nevertheless, as we will see, in the case of the two-point function it comprises no major difficulty and leads to the correct result.

\subsection{Integration of the zero-mode}

Let us first integrate the zero mode of $X^0 (z)$. In order to do so, let us split the field in its expectation value $\langle X^0 \rangle = x^0$ and its fluctuations $\tilde{X}^0(z)$ around it; namely, we make $X^0 (z)= \tilde{X}^0(z) +x^0$, with $\langle \tilde{X}^0 \rangle =0$. This results in 
\begin{eqnarray}
\mathcal{A}_2' (p^0_1)&=& \int_{\tilde{X}^0_{(\mathbb{CP}^1)}} \frac{\mathcal{D}\tilde{X}^0 \, }{\text{Vol}{(Res)}} e^{\frac{1}{4\pi}\int_{\mathbb{CP}^1} [(\partial \tilde{X}^0)^2-\frac{Q}{\sqrt{2}} R\tilde{X}^0 ]}
e^{\sqrt{2}\alpha \tilde{X}^0(0)}\, e^{\sqrt{2}\alpha \tilde{X}^0(1)}\times \nonumber \\
&& \int _{\mathbb{R}} dx^0  \,   e^{\sqrt{2}x^0(2\alpha -\frac{Q}{8\pi } \int_{\mathbb{CP}^1} R )} 
e^{-\mu \int_{\mathbb{C}} \exp (\sqrt{2}bX^0)} \label{Indi}
\end{eqnarray}%
which we can write in the following form
\begin{eqnarray}
\mathcal{A}_2' (p^0_1)&=&\int _{\tilde{X}^0_{(\mathbb{CP}^1)} } \frac{\mathcal{D}\tilde{X}^0 \, \, e^{-S_L[\mu =0 ]}}{\text{Vol}{(Res)}}\, e^{\sqrt{2}\alpha \tilde{X}^0(0)}\, e^{\sqrt{2}\alpha \tilde{X}^0(1)}\times \nonumber \\
&& \int _{\mathbb{R}} dx^0 \, \int _{\mathbb{R_{+}}} d\eta  \,   e^{\sqrt{2}x^0(2\alpha -\frac{Q}{8\pi } \int_{\mathbb{CP}^1} R )} e^{-\mu \eta } \, \delta \left(\eta -e^{\sqrt{2}bx^0 }\int_{\mathbb{C}}e^{\sqrt{2}b\tilde{X}^0 } \right) 
\end{eqnarray}%
by introducing a $\delta$-function. The integral in $\eta$ goes over $\mathbb{R}_{\geq 0}$ because the marginal operator $\int_{\mathbb{C}} d^2z \exp\Big({\sqrt{2}bX^0 (z)}\Big)$ has bounded support. Now, we can use basic properties of the $\delta $-function\footnote{i.e. the composition of the linear functionals, $\delta(f(x))=\sum_{\{x_i/f(x_i)=0\}} \delta(x-x_i)/f'(x_i)$.} to write
\begin{eqnarray}
\mathcal{A}_2' (p^0_1)&=&\int _{\tilde{X}^0_{(\mathbb{CP}^1)} } \frac{\mathcal{D}\tilde{X}^0 \, \, e^{-S_L[\mu =0 ]}}{\text{Vol}{(Res)}}\, e^{\sqrt{2}\alpha \tilde{X}^0(0)}\, e^{\sqrt{2}\alpha \tilde{X}^0(1)}\times \nonumber \\
&& \int _{\mathbb{R}}
dx^0 \, \int _{\mathbb{R_{+}}} d\eta  \,  \, e^{\sqrt{2}x^0(2\alpha -\frac{Q}{2 }  \chi{ (\mathbb{CP}^1)})}  \,   \delta \left(x^0  +\frac{1}{\sqrt{2}b} \log (\eta^{-1}
\int_{\mathbb{C}}e^{\sqrt{2}b\tilde{X}^0 })
\right) 
\end{eqnarray}%
where the Euler characteristic of the sphere can be computed using the Gauss-Bonnet theorem, namely $\frac{1}{4\pi }\int _{\mathbb{CP}^1}R=\chi({S^2})=2$. Commuting integrals and integrating over $x^0$, we get\footnote{We absorb a factor $\sqrt{2}$ in the definition of the measure.}
\begin{eqnarray}
\mathcal{A}_2' (p^0_1)&=& \int _{\tilde{X}^0_{(\mathbb{CP}^1)} } \frac{\mathcal{D}\tilde{X}^0 \, \, e^{-S_L [\mu =0 ]}}{b\, \text{Vol}{(Res)}} \, e^{\sqrt{2}\alpha \tilde{X}^0(0)}\, e^{\sqrt{2}\alpha \tilde{X}^0(1)}\,  \left( \int_{\mathbb{C}}e^{\sqrt{2}b\tilde{X}^0 } \right)^{s}  \int _{\mathbb{R_{+}}}  d\eta\, \eta^{-1-s}   {e^{-\mu \eta }}\label{tormentita}
  \end{eqnarray}
where $s=(Q-2\alpha )/b$. The last factor in this expression produces a factor $\mu^s\Gamma (-s)$ together with the insertion of $s$ marginal operators $\int d^2w\, \exp{(\sqrt{2}bX^0(w))}$. This is the standard trick in the Coulomb gas realization of a non-compact CFT, cf. \cite{GL}. This means that expression (\ref{tormentita}) has actually to be written in its operatorial form; namely
\begin{equation}
\mathcal{A}_2' (p^0_1)=\frac{\mu ^s\Gamma(-s)}{b}\, \int_{\mathbb{C}^s}\prod_{r=1}^sd^2w_r\, \int _{\tilde{X}^0_{(\mathbb{CP}^1)} } \frac{\mathcal{D}\tilde{X}^0 \, \, e^{-S_L[\mu =0 ]}}{\text{Vol}{(Res)}}\, e^{\sqrt{2}\alpha_{(1)}\tilde{X}^0(0)}\, e^{\sqrt{2}\alpha_{(1)}\tilde{X}^0(1)}\,   \prod_{l=1}^s   e^{\sqrt{2}b\tilde{X}^0 (w_l) } \, .
  \label{nwat}\end{equation}
It is obvious that this expression is merely formal unless $s=(Q-2\alpha )/b=-2ip^0_1/b$ is a positive integer number. In such case, and in the limit $b\to 1$ we are interested in, the energy spectrum in the Euclidean theory turns out to be $ ip^0 \in \frac 12\mathbb{Z}_{<0}$. However, the physical region corresponds to complex $s$, for which (\ref{tormentita}) can only be taken {\it formally}. Therefore, for generic values of $s$ an analytic extension of the formula (\ref{tormentita}) is needed. Such extension amounts to first solve the integrals for $s\in \mathbb{Z}_{\geq 0}$ and then extend the resulting expression to $s\in \mathbb{C}$. Such an extension, however, is not uniquely defined as it is an extension of an expression originally defined over $\mathbb{Z}_{>0}$ to an expression defined over $\mathbb{C}$. In particular, phase ambiguities arise. Nevertheless, this procedure happens to be under control and can be carried out in a very natural way: As a matter of fact, the integrals involved yield expressions in terms of quotients and products of $\Gamma $-functions and then, at the end, it reduces to analytically extending combinatorial expressions. 

\subsection{Interpretation of (\ref{nwat})}

Formula (\ref{nwat}) corresponds to an $s$-point function in the timelike free boson theory $S[\mu = 0]$. The insertion of the $s$ integrated operators $\int d^2z\,\exp (\sqrt{2}bX^0 (z))$ have a natural interpretation: from the conformal field theory point of view, they are a precise amount of screening operators needed to screen the background charge $Q$ at infinity. The presence of integrated marginal operators also admits a natural target space interpretation: they can be thought of as the tachyons that constitute the exponential wall; these tachyons interact among themselves and also with the two vertices. From string theory perspective, Liouville field theory coupled to additional matter (free bosons $X^{i}(z)$ with $i =1, 2, ... , D-1$) can actually be thought of as a $\sigma$-model describing a tachyon-dilaton background with $D-1$ flat directions and one non-compact $X^0$-dependent direction. The fact that the $s$ insertions are integrated means that they are hard sources that constitute the wall, and so they are integrated over the whole non-compact direction $X^0$. According to this, the prefactor $\Gamma(-s)$ in (\ref{nwat}) has the same interpretation as the one given in \cite{diFK}; it is the factor associated to the poles of resonant correlators with a definite number of tachyon insertions.

\subsection{Alternative derivation of (\ref{nwat})}

An alternative way of arriving to expression (\ref{nwat}) is to go back to expression (\ref{Indi}) and, there, expand the interaction term as follows
\begin{eqnarray}
e^{\sqrt 2x^0(2 \alpha -\frac{Q}{{2}}\chi{ (\mathbb{CP}^1)})}\, e^{-\mu \int_{\mathbb{C}}d^2w \exp(\sqrt{2}b X^0 (w))}=e^{\sqrt{2}(2\alpha -Q)x^0}\,\sum_{s=0}^{\infty }\frac{(-1)^s\mu^s}{s!} e^{\sqrt{2}bx^0 s}\,\int_{\mathbb{C}^s}  \prod_{l=1}^s\left( d^2w_l\,  e^{\sqrt{2}b \tilde{X}^0 (w_l)} \right)\nonumber
\end{eqnarray}
Then, we can integrate over the zero mode $x^0$, which only appears in exponentials. This generates a $\delta$-function factor; namely
\begin{eqnarray}
&&\sum_{s=0}^{\infty }\frac{(-1)^s\mu^s}{\sqrt{2}\, s!}\, \delta \left(2\alpha +b s  - Q \right) \int_{\mathbb{C}^s}  \prod_{l=1}^s\left( d^2w_l\,  e^{\sqrt{2}b \tilde{X}^0 (w_l)} \right) =  \nonumber \\
 &&\ 
 \ \ \ \ \frac{(-\mu )^{(Q-2\alpha )/b} \, \delta (0)}{\sqrt{2}\, ((Q-2\alpha )/b)!}   \,  \int_{\mathbb{C}^s}  \prod_{l=1}^{(Q-2\alpha)/b}\left( d^2w_l\,  e^{\sqrt{2}b \tilde{X}^0 (w_l)} \right)
\end{eqnarray}
The evaluation of this $\delta$-function produces a divergent factor $\delta(0)$, which we wrote above in an informal way but which can be treated more carefully; for example, by writing
\begin{eqnarray}
\lim_{s\to \frac{Q-2\alpha }{b}}\frac{(-1)^s\mu^s}{s!}\delta(Q -b s-2\alpha )&=& 
 \lim_{\varepsilon\to 0}\frac{(-\mu )^{\frac{Q-2\alpha }{b}}\Gamma(\varepsilon )}{\Gamma(1+(Q-2\alpha )/b)}
 = \mu^{\frac{Q-2\alpha }{b}}\lim_{\varepsilon \to 0}\Gamma((2\alpha -Q)/b
 -\varepsilon).\nonumber
\end{eqnarray}
This reproduces the prefactor in (\ref{nwat}).

\subsection{Fixing the Killing conformal group}

Now, let us deal with the volume of the conformal Killing group, i.e. the stabilizer of $PSL(2,\mathbb{C})$. The volume of this non-compact group can be parameterized as follows
\begin{equation}
\text{Vol}_{(SL(2,\mathbb{C}))} = \int_{\mathbb{C}^3}\frac{d^2z_1\, d^2z_2\,d^2w_1}{|z_1-z_2|^2|z_2-w_1|^2|w_1-z_1|^2} \, ,\label{quince}
\end{equation}
which is standard in the string theory amplitude computations. This permits to cancel the factor (\ref{quince}) in the denominator of (\ref{nwat}) by simply fixing the two vertex operators and, along with them, one of the $s$ marginal operators. As usual, we choose $z_1=0$, $z_2=1$, and $w_1=\infty $. The latter insertion has to be understood as being accompanied with the appropriate factor that extract the singularity, namely in the $\lim_{w_{1}\rightarrow \infty
}|w_{1}|^{4} e^{\sqrt{2}bX^0
(w_{1})}$. This yields
\begin{equation}
\mathcal{A}_2' (p^0_1)=\frac{\mu ^{s}}{b}\Gamma (-s)
\int_{\mathbb{C}^{s-1}} \prod_{t=2}^{s} d^{2}w_{t}\int_{\tilde{X}^0_{(\mathbb{CP}^1)}} \mathcal{D}\tilde{X}^0 \, e^{-S_L[\mu = 0]}\, e^{\sqrt{2}\alpha \tilde{X}^0 (0)}e^{\sqrt{2}\alpha \tilde{X}^0 (1)}e^{\sqrt{2}b\tilde{X}^0
(\infty )}\prod_{r=2}^{s}e^{\sqrt{2}b\tilde{X}^0 (w_{r})}  \label{empieza}
\end{equation}
with $s=(Q-2\alpha )/b$. This is exactly the point where the divergence coming from the residual conformal symmetry is cancelled out. By fixing one screening operator at $w_1=\infty $, we manifestly see how the time-dependent background happens to stabilize the worldsheet residual conformal symmetry. 

In this way, we are left with an $(s+2)$-point correlation function of a free theory; namely
\begin{equation}
\mathcal{A}_2' (p^0_1)=\frac{\mu ^{m+1}}{b}\Gamma (-m-1)
\int_{\mathbb{C}^m} \prod_{t=1}^{m} d^{2}w_{t}\,
\Big\langle  e^{\sqrt{2}\alpha  \tilde{X}^0 (0)}e^{\sqrt{2}\alpha \tilde{X}^0 (1)}e^{\sqrt{2}b\tilde{X}^0
(\infty )}\prod_{r=1}^{m}e^{\sqrt{2}b\tilde{X}^0 (w_{r})} \Big\rangle_{\text{free}}  \label{empiezaggg}
\end{equation}
where $m=s-1=(Q-2\alpha )/b-1=-2ip^0_1/b-1$. Therefore, we can try to solve it by using free field techniques. However, before doing so, we have to take care of the following issue: As pointed out before, expressions like (\ref{empieza}) are only well-defined for $s\in \mathbb{Z}_{>1}$. In order to make sense out of such a formula for generic values of $s$, we have perform an analytic continuation. In order to do so, we consider the multiple integral expression coming from (\ref{empiezaggg}), we solve it for $m=s-1=-2ip^0_1/b-1\in \mathbb{Z}_{\geq 1}$, we write the combinatoric factors in terms of quotients of $\Gamma$-functions, and then we analytically extend the final expression. We do this in detail in the following subsection.

\subsection{Coulomb gas and conformal integrals}

The correlators in the timelike free theory on $\mathbb{CP}^1$ can be computed using the Green function
\begin{equation}
\Big\langle\tilde{X}^0 (w_i) \tilde{X}^0 (w_j)\Big\rangle_{\text{free}} = + \log |w_i - w_j|^2\, ,
\end{equation}
which implies
\begin{equation}
\langle e^{\sqrt{2}\alpha\tilde{X}^0 (w_i)} e^{\sqrt{2}b\tilde{X}^0 (w_j)}\rangle_{\text{free}} = e^{ 2\alpha b\Big\langle\tilde{X}^0 (w_i) \tilde{X}^0 (w_j)\Big\rangle_{\text{free}}}\,=\, |w_i - w_j|^{4\alpha b}\, .
\end{equation}
Then, applying the Wick theorem, we arrive to the following integral expression
\begin{equation*}
\mathcal{A}_2' (p^0_1)=b^{-1}\mu ^{m+1}\Gamma (-m-1)\int_{\mathbb{C}^m} \prod_{i=1}^{m} d^{2}w_{i}\, 
\prod_{r=1}^{m}|w_{r}|^{4\alpha b }|1-w_{r}|^{4\alpha b
}
\prod_{t=1}^{m}
\prod_{\ell =1 }^{t-1}
|w_{t}-w_{\ell }|^{4 b^2 } ,
\end{equation*}%
which is actually an extension of Shapiro-Virasoro integral. This integral has been solved by Dotsenko and Fateev in \cite{Fateev2}. Assuming $m\in \mathbb{Z}_{\geq 1}$, this yields\footnote{We absorb a factor $\pi $ in the definition of the amplitude.}
\begin{equation}
\mathcal{A}_2' (p^0_1)=\frac{(\pi \mu )^{m+1}}{ b}\Gamma (-m-1)\Gamma (m+1)\,\gamma
^{-m}(b^2 )\prod_{r=1}^{m}\frac {\gamma (rb^2 )\gamma
^{2}(1+2\alpha b+(r-1)b^2 )}{\gamma (2+4\alpha b +(m-2+r)b^2 )}\, ,\label{Postalina}
\end{equation}
where $\gamma (x)=\Gamma (x)\Gamma ^{-1}(1-x)$. It is convenient to rewrite this expression using functional properties of the $\gamma $-function and momentarily assuming $m\in \mathbb{Z}_{>0}$.  Among the properties of the $\gamma$-function we will need, there are the reflection property $\gamma(1-x)=\gamma^{-1}(x)$ and the shift property and $\gamma(1+x) = -x^2\gamma(x)$. Using this and the fact $m=s-1=-b^{-2}-2\alpha b^{-1}$, we can rewrite one of the factors in (\ref{Postalina}) as follows
\begin{equation}
\prod_{r=1}^{m} \gamma (rb^2) \gamma (1+2\alpha  b +(r-1)b^2) = \prod_{r=1}^{m} \gamma (rb^2) \gamma (-rb^2) =
\frac{(-1)^{s-1}}{b^{4(s-1)}}\frac{1}{\Gamma^2 (s)}\, . \label{HJ1}
\end{equation}
In a similar way, we can rewrite another factor as follows
\begin{equation}
\prod_{r=1}^{m} \frac{ \gamma (1+2\alpha  b +(r-1)b^2)}{\gamma(2+4\alpha  b +(m-2+r)b^2)} =\frac{\gamma (b^2-2\alpha  b)}{\gamma (1+b^2)} \, .\label{HJ2}
\end{equation}
Replacing (\ref{HJ1}) and (\ref{HJ2}) in (\ref{Postalina}) and, again, using some properties of the $\gamma$-function, we finally obtain the following result
\begin{equation}
\mathcal{A}_2' (p^0_{(1)})=\left( \pi \mu \gamma (-b^2)\right)^{\frac{Q-2\alpha }{b}} \frac{\gamma (b^{-2}+2b^{-1}\alpha  ) \gamma (b^{2}-2b\alpha  )}{(2\alpha -b+b^{-1}  )}\, , \label{Postalinz}
\end{equation}%
which turns out to be the timelike Liouville two-point function, with $\alpha = Q/2+ip^0_{(1)}$.

\subsection{Restoring Poincar\'e symmetry}

Now, we can take the limit (\ref{LaLimite}) of the expression (\ref{Postalinz}), which actually corresponds to removing the Liouville deformation and {\it ipso facto} restoring time translation symmetry. Replacing $b=1+\varepsilon $ in (\ref{Postalinz}), taking the limit $\varepsilon \to 0$, which implies $Q\simeq 2\epsilon$, and using basic properties of the $\Gamma $-function, such as
\begin{equation}
\frac{\Gamma (1+2ip^0_{(1)})\Gamma (1-2ip^0_{(1)})}{\Gamma (2ip^0_{(1)})\Gamma (-2ip^0_{(1)})} \, = \, (2p^0_{(1)})^2 \, ,
\end{equation}
we find
\begin{equation}
\lim _{\varepsilon \to 0}\mathcal{A}_2' (p^0_1)\, =\, {2 p_{(1)}^0} \lim _{\varepsilon \to 0}\left( \frac{\mu  \pi }{2\varepsilon }\right) ^{(\varepsilon +1)(p^0_{(1)}+p^0_{(2)})} 
\end{equation}
which here we wrote in terms of the Euclidean momentum $p^0_{(1)}\to i p^0_{(1)}$, i.e. $\alpha = -p_{(1)}^0=-p_{(2)}^0$. This expression manifestly shows that the choice (\ref{LaLimite2}) was the correct one. On the one hand, it corresponds to turning off the Liouville potential by keeping the KPZ scaling finite; on the other hand, it renormalizes the Liouville cosmological constant $\mu $ as usual in the $c_L\to 1$ limit of Liouville theory, cf. \cite{S}. In fact, we can chose $\mu_* = -Q/\pi$ in (\ref{Diego}) to finally obtain 
\begin{equation}
\mathcal{A}_2(p^{\mu}_{(1)},p^{\mu}_{(2)}) =   2p_{(1)}^0\, (2\pi )^{D-1}\delta^{(D-1)} (p^i_{(1)}+ p^i_{(2)})\, ;\label{SP2eeee}
\end{equation}
with the $\delta $-function in the momenta coming from the integration over the zero-modes of the spacelike fields, $x^i$. Equation (\ref{SP2eeee}) exactly reproduces the free particle expression (\ref{SP2eeee}); cf. (2.15) in \cite{Malda}, (2.51) in \cite{190903672}, and (15) in \cite{2012.03802}. Notice that a different choice for the numerical coefficient $\kappa \equiv \pi\mu_*/Q$ would have been absorbed in the normalization of the vertices, as $V_{p_{(a)}}\to \kappa^{p_{(a)}^0} V_{p_{(a)}}$; this is because the KPZ scaling of the Liouville $N$-point functions in the limit $Q\to 0$ is given by $\kappa ^{-\sum_{a=0}^Np_{(a)}^0}$. 

\section{Concluding remarks}

By considering the Liouville dimension as a regularizing parameter, we have obtained the correct expression for the string two-point amplitude (\ref{SP2eeee}), including the correct dependence with the energy, $2p^0_{(1)}$. This is consistent with the normalization in (\ref{ecua}); notice that the latter equation can be written as
\begin{equation}
\mathcal{A}_2(p^{\mu}_{(1)},p^{\mu}_{(2)}) = \int_{\mathbb{R}^{D}}\frac{d^{D-1}q}{(2\pi )^{D-1}}\frac{1}{2q^0}\,
\mathcal{A}_2(q^{\mu} , p^{\mu}_{(1)})\, \mathcal{A}_2(q^{\mu},p^{\mu}_{(2)}) \, . \label{ecuaz} 
\end{equation}
The crucial ingredient in our calculation was the Liouville two-point correlation function; namely $\langle  e^{\sqrt{2}\alpha X(z_1)} e^{\sqrt{2}\alpha X(z_2)}  \rangle_{L}= |z_1 -  z_2|^{4\alpha (\alpha -Q)}\, B_L(\alpha )$, with $X(z)$ being the Liouville field. In relation to the dependence of the two-point amplitude (\ref{SP2eeee}) with the energy, it is worth not to mistake the Liouville two-point function for the Liouville reflection coefficient $R_L (\alpha )$. While both quantities are closely related, they differ by a factors that is crucial for our computation; namely ${R_L (\alpha )}/{B_L (\alpha )}= {\pi}/{(Q-2\alpha )}$; see (4.20)-(4.21) in \cite{perturbed}. In fact, if instead of considering the two-point function we started with the reflection coefficient, then we would had never gotten the proper factor $2p_{(1)}^0$ in (\ref{SP2eeee}), but rather 
\begin{equation}
\lim _{b\to 1}R_L (\alpha ) = -\frac{\pi }{2p _{(1)}^{0}} \lim _{\varepsilon\to 0} \mathcal{A}'(p_{(1)}^0) \, ; 
\end{equation}
recall that $\alpha ={Q}/{2}+ip_{(1)}^0$. About the factor $2p^0_{(1)}$, see the discussion around equation (4.13) in \cite{Becker}, and see also equations (3.3) and (A.6) in \cite{MO3}, and equations (4.16) and (4.49) in \cite{Pacman}. --All those references discuss a similar factor appearing in the two-point function of the $SL(2,\mathbb{R})$ WZW model, which is closely related to the Liouville two-point function \cite{Ribault}.-- Also in relation to the reflection coefficient in the timelike theory, it would be interesting to think whether one could interpret our calculation as that of the particle creation rate in the time-dependent background, whose analytic continuation gives the correct result for the two-point function.

As a consistency check of expression (\ref{Postalinz}) for the timelike Liouville two-point function, we can observe that this corresponds to the analytic extension $b\to ib$ of the spacelike Liouville two-point function. Also, it is related to the timelike partition function ${Z_L}$ as the analogous spacelike observables are; namely, we have
\begin{equation}
\lim_{p^0_{(1)}\to iQ/2} \mathcal{A}_2' (p^0_{(1)}) = \frac{\pi^2}{1+b^2}\, {Z_L}\, ;
\end{equation}
cf. \cite{perturbed}; $\alpha = Q/2+ip_{(1)}^0$. It would be interesting to explore an adaptation of our techniques to the computation of the partition function. 

Another important remark is about the three-point function: Actually, the question might arise as to whether we are not being too naive when thinking that the limit $b\to 1$ of the timelike Liouville theory is actually equivalent to recovering the time-translation invariant theory. As a matter of fact, it is a known result that such $c_L\to 1$ limit is subtle, especially regarding the tree-point function, and that $c=1$ interacting conformal field theories that obey consistency conditions are obtained in such limit \cite{190512689, Rodriguez}. In \cite{S}, the author showed that two possible non-trivial $c=1$ theories appear as a limit from Liouville theory; this depends on the possible values of the Liouville momenta $\alpha_{(a)}$: while for purely imaginary momenta the limiting theory is identified with the theory proposed in \cite{RW}, which is itself a limit of unitary minimal models, for other values of $\alpha_{(a)}$ a different theory is obtained, cf. \cite{ST}. The timelike $c=1$ three-point function is known to exhibit special features. In particular, the limiting procedure connecting the three-point function to the two-point function is obstructed due to the existence of a dimension-$0$ operator, other than the identity, in the timelike theory \cite{HMW, 190512689, Saleur}. The special features of the timelike three-point function were discussed by many authors, and notably by Harlow, Maltz, and Witten in \cite{HMW}; see also \cite{Giribet}-\cite{Mati} and references therein and thereof. Our computation, however, only relies on the expression for the two-point function, for which no special features are expected to appear and the result was expected to yield the correct two-point function as shown. 


\[\]

G.G. is indebted to Mauricio Leston for discussions. The work of N.L. has been partially supported by the Dean’s Undergraduate Research Fund Grant from FAS-NYU.

\end{document}